\documentclass[12pt,preprint]{aastex}

\newcommand{\be}{\begin{equation}}
\newcommand{\ee}{\end{equation}} 
\newcommand{\simless}{\lower.5ex\hbox{$\; \buildrel < \over \sim\;$}}
\newcommand{\simgreat}{\lower.5ex\hbox{$\; \buildrel > \over \sim\;$}}

\begin{document}

\title{Cosmic Ray Propagation in Turbulent Spiral Magnetic Fields \\
associated with Young Stellar Objects} 

\author{Marco Fatuzzo$^1$ and Fred C. Adams$^{2,3}$} 

\affil{$^1$Physics Department, Xavier University, Cincinnati, OH 45207}
\affil{$^2$Physics Department, University of Michigan, Ann Arbor, MI 48109}
\affil{$^3$Astronomy Department, University of Michigan, Ann Arbor, MI 48109}

\begin{abstract}

External cosmic rays impinging upon circumstellar disks associated
with young stellar objects provide an important source of ionization,
and as such, play an important role in disk evolution and planet
formation.  However, these incoming cosmic rays are affected by a
variety of physical processes internal to stellar/disk systems,
including modulation by turbulent magnetic fields.  Globally, these
fields naturally provide both a funneling effect, where cosmic rays
from larger volumes are focused into the disk region, and a magnetic
mirroring effect, where cosmic rays are repelled due to the increasing
field strength. This paper considers cosmic ray propagation in the
presence of a turbulent spiral magnetic field, analogous to that
produced by the Solar wind. The interaction of this wind with the
interstellar medium defines a transition radius, analogous to the
Heliopause, which provides the outer boundary to this problem.  We
construct a new coordinate system where one coordinate follows the
spiral magnetic field lines and consider magnetic perturbations to the
field in the perpendicular directions.  The presence of magnetic
turbulence replaces the mirroring points with a distribution of values
and moves the mean location outward. Our results thus help quantify
the degree to which cosmic ray fluxes are reduced in circumstellar
disks by the presence of magnetic field structures that are shaped by
stellar winds.  The new coordinate system constructed herein should
also be useful in other astronomical applications.
 
\end{abstract} 

\keywords{Cosmic Rays -- diffusion -- ISM -- molecular clouds}

\section{Introduction}
\label{sec:intro} 

Circumstellar disks provide the birth places for planets and the
corresponding disk properties constrain the possible architectures of
the resulting planetary systems. Moreover, most of the mass that
eventually becomes incorporated into a forming star falls initially
onto the accompanying disk.  The formation, evolution, and eventual
demise of circumstellar disks thus represents a fundamental issue in
star and planet formation. Ionization plays an important role in
setting the thermal (Glassgold \& Langer 1973), dynamical (Balbus \&
Hawley 1991; Gammie 1996), and chemical (Semenov et al. 2004; Cleeves
et al. 2013) properties of these planet-forming disks. The ionization
levels in young stellar objects are determined by many sources,
including cosmic rays, short-lived radioactive nuclei, and photons of
sufficiently high energy. Each of these sources can be important and
will contribute to the ionization levels in different parts of the
disk. In addition, each of these sources must be studied in detail to
understand their roles, but such a treatment is beyond the scope of a
single paper. This paper focuses on the propagation of cosmic rays
\citep{hayakawa,spitzer} in the turbulent magnetic fields that surround
young stellar objects.

Magnetic fields have a significant influence on the propagation of
cosmic rays.  Large-scale field structures can mirror and focus
charged particles, whereas resonant coupling between charged particles
and small-scale fluctuations on the order of the particle gyration radius leads to diffusive
motion. The gyration radius therefore determines the length scale that divides
these two regimes.  As shown in Sections 4 and 5, the radius of gyration 
for protons with Lorentz factors $\gamma \la 10^3$ is smaller than the 
length scale over which the large-scale field structure changes appreciably
within the T Tauriosphere.
Moreover, we expect
these environments to be turbulent over a range of wavelengths 
that spans over the gyration radius for most galactic cosmic rays. 
As such, both reflection and diffusion effects come into play in the region
immediately surrounding young star/disk systems for the great majority
of galactic cosmic-rays.  

Magnetic field structures are expected to vary with the evolutionary
stage of the objects. During the early phases, as gravitational
collapse leads to disk formation, magnetic field lines are dragged
inward to produce magnetic field structures with an hour-glass form.
This geometry arises both in the limit where gravity overwhelms the
field during collapse (see, e.g., \citealt{padoan,maclow}), 
and in the limit where the flow is magnetically
controlled (\citealt{shu,galli,ostriker}). 
Observations of forming stars indicate the presence of
such hour-glass-like magnetic fields (\citealt{davidson,qiu}) and find
approximate alignment between the background magnetic fields and the
symmetry axis of the infalling envelopes (Chapmann et al. 2013).  During later
phases of evolution, stellar winds and outflows sculpt the magnetic
field geometry.  In the simplest picture, the rotating stellar wind
pushes out a magnetic field structure with a simple spiral form 
(Parker 1958). In both limits, turbulence produces magnetic field
fluctuations that are superimposed on the large-scale field
structures. These fluctuations, in turn, affect the propagation of
cosmic rays into the disk and influence their efficacy as an
ionization source. In an earlier paper (Fatuzzo \& Adams 2014), we considered
cosmic ray propagation in hour-glass field configurations expected in
the earliest stages of protostellar evolution.  In this paper, we
consider cosmic ray propagation in the later stages when the magnetic
field has a spiral form, with additional magnetic field fluctuations
due to turbulence.

Previous work has considered the propagation of cosmic rays in young
stellar objects, including both funneling and mirroring effects (e.g.,
Padoan \& Scalo 2005; Desch et al. 2004; Padovani \& Galli 2011).
This work shows that mirroring
effects tend to dominate over focusing effects, thereby reducing the
net ionization rate by a factor $f\sim2-3$. Additional suppression of
the cosmic ray flux can result from the twisting of magnetic field
lines during protostellar collapse (Padovani et al. 2013). 
Further suppression can occur due to turbulent fluctuations in 
the magnetic field lines (Fatuzzo \& Adams 2014), as considered herein. 

For revealed star/disk systems, another mechanism arises for the
suppression of incoming cosmic rays.  The central stars (T Tauri
stars) are often observed to have strong stellar winds, which are
generally more powerful than that of the Sun. As is well known, the
Solar wind, which follows the spiral structure out to the radius where
it interacts with the interstellar medium, acts as a barrier to cosmic
rays.  The inner boundary of this interaction region is known as the
termination shock, and lies at $\sim90$ AU in our Solar System. The
outer boundary of this interaction region marks the boundary where the
wind is stopped by the interstellar medium (at a distance of $\sim120$
AU).  For the stronger winds associated with T Tauri stars, the
interaction region -- the T Tauriosphere -- will extend farther out,
with an expected size of $\sim1000$ AU for typical systems (Cleeves et
al. 2015).  Incoming cosmic rays will thus be attenuated as they try
to penetrate into the T Tauriosphere. The boundary of this region --
essentially the termination shock of the T Tauriosphere -- marks the
outer boundary of the problem considered here. Within this boundary,
the wind from the young star is expected to follow the spiral form
analogous to the Parker spiral, as considered in this paper.

One important aspect of this paper is thus to include the effects of
magnetic turbulence on the propagation of cosmic rays through magnetic
field configurations. To accomplish this goal, we construct a novel
(non-standard) coordinate system where one coordinate follows the
magnetic field lines. This approach facilitates the analysis by
allowing for a straightforward implementation of the field
fluctuations subject to the required constraint $\nabla\cdot{\bf B}=0$
(see also Fatuzzo \& Adams 2014). In other words, we can easily define
magnetic field perturbations that point in the orthogonal directions
and remain divergence-free. This coordinate system, and the methods 
used to construct it, should be useful
for other astronomical applications, in addition to the study of 
cosmic ray propagation carried out here (e.g., Adams 2011;  Adams \& Gregory 2012). 

In the adopted magnetic field geometry, cosmic-rays that spiral inward
along the equatorial plane would have the greatest path-length through
the disk, and at face value, the greatest potential for ionizing the
disk.  However, for the expected T-Tauri disk environments, the
stopping length of cosmic-rays at the mid-plane of the outer edge of
the disk is found to be $L\sim 35$ AU (see last paragraph in Section 5), indicating that
ionization in this region is limited to the outer part of the disk.
The biggest impact on the overall disk is therefore expected to result
from particles coming inward near the equatorial plane, but moving a
few scale-heights above.  In this paper, we focus primarily on the
effect that turbulence has on cosmic-ray propagation along the
equatorial plane, and use the results to draw general conclusions
about the overall (global) effect on disk ionization.  A complete
analysis of the ionization structure of T-Tauri disks will then be
left for future work. 

This paper is organized as follows. We first construct a new
coordinate system that follows the spiral magnetic field lines in
Section 2.  Turbulent fluctuations are introduced in Section 3, where we use the
new spiral coordinate system to enforce divergence-free perturbations
in the magnetic fields.  We consider the general motion of charged particles
through the spiral field in Section 4, and explore the magnetic mirroring problem
in the presence of magnetic turbulence in Section 5 to find how
the ensuing distributions of reflection points for different field profiles 
compare to their classical (turbulence-free) counterparts.
The paper concludes in Section 6 with a
summary of our results and a discussion of their implications.

\section{Parker Spiral Magnetic Field Geometry} 

This section first presents the Parker spiral magnetic field geometry,
and then constructs a new co-ordinate system that naturally represents 
that geometry. In spherical coordinates ($\xi,\theta,\phi$), the
equation for the field lines of a Parker spiral takes the form 
\be
\xi-1-\ln \xi = {v_w\over \omega R_*} \phi \sin\theta\,,
\label{spiral} 
\ee 
where $R_*$ is the stellar radius, $v_w$ is the stellar wind speed
(typically in the range 200 -- 400 km/s, with the higher value
representative of our Sun), $\omega = v_{rot} / R_*$ is the angular
speed at the stellar surface, and $\xi \equiv r/R_*$ provides a
convenient way to express radial distance in a non-dimensional form.
Observations yield measured values of $v_{rot} \approx 1.8$ km/s for
our Sun, and significantly greater values for T Tauri stars (where
$v_{rot}$ is expected to fall in the range $10-100$ km/s). Likewise,
T Tauri stars have radii that are larger than our Sun by about a
factor of 2 (so that $R_* \sim 1 - 2 \times 10^{11}$ cm). These physical 
parameters can be combined into a single ``field structure" parameter 
$A$ defined by 
\be
A\equiv {v_w\over \omega R_*} = {v_w \over v_{rot}} \,. 
\ee
For T Tauri stars, this parameter is expected to fall in the range
$A=2-40$ (compared to $A \approx 200$ for the Sun).

With the condition that field lines spiral outward along cones of
constant angle $\theta_0$, equation (\ref{spiral}) can be used to
define a new coordinate 
\be
q \equiv \xi - 1 - \ln{\xi} - A \phi  \sin \theta_0 \,,
\ee
that remains constant along a magnetic field line.  
The magnetic field, in turn, must be perpendicular to the gradient
\be
\nabla q = {\xi-1\over\xi}\hat\xi - {A \sin\theta_0 \over\xi \sin\theta} \hat\phi \,.
\ee
With the further requirements that $\theta = \theta_0$ and 
$\nabla\cdot {\bf B}_P = 0$, one then readily obtains the expression
\be
{\bf B}_P = B_* \left[ {\hat \xi \over \xi^2} + {\xi - 1 \over A \xi^2} \hat\phi \right]\,,
\ee
where $B_*$ is the magnetic field strength on the stellar surface ($\xi = 1$).
We note that the magnitude of the field is then given by 
\be
B_P = B_* {\sqrt{A^2 + (\xi - 1)^2}\over A \xi^2}\,,
\ee
and $B_P  \rightarrow \xi^{-1}$ as $\xi \rightarrow\infty$.  

Adopting the formalism of Fatuzzo \& Adams (2014), we treat $q$ as one
of the coordinates in an orthogonal coordinate system that can be used
to naturally represent the spiral magnetic field structure.  A second
coordinate $p$ that also exists on a constant $\theta_0$ cone is
constructed by noting that $\nabla p$ must be parallel to $\hat B_P$
(and hence perpendicular to $\nabla q$).  As shown in Appendix A, an
infinite number of solutions exist.  We adopt here the only one that
neither diverges or converges exponentially with $\xi$: 
\be
p = A \ln \left(1-{1\over\xi}\right) + \phi  \sin\theta_0\,.
\ee
We note, however, that $p \rightarrow - \infty$ in the limit $\xi
\rightarrow 1$, making it impossible to ``connect" a line of constant
$p$ back to the surface.  The third coordinate is found through the
cross-product $\nabla p \times \nabla q$, which points in the $\hat
\theta$ direction.  Without loss of generality, we can then use
$\theta$ as the third coordinate\footnote{Notice that since $\xi$ and
  $\phi$ remain constant along the $\theta$ axis, both $p$ and $q$
  must then change.  In essence, our construction of a coordinate
  system around a cone of constant $\theta$ should therefore be viewed
  as a local treatment.}, which then also remains constant as one
moves along a field line.  We have thus defined a new coordinate
system ($p,q,\theta$) in terms of the original coordinates
($\xi,\theta,\phi$).

The dimensionless covariant basis vectors $\underline{\epsilon}_j$ are 
given by the usual relations
\be
\underline{\epsilon}_p =  \,\nabla p, \qquad \underline{\epsilon}_q = \,\nabla q, 
\qquad {\rm and} \qquad \underline{\epsilon}_\theta = \,\nabla \theta\,,
\ee
where the gradient is written in terms of the variables ($\xi$, $\theta$, $\phi$).
We note that the quantities $\underline{\epsilon}_j$ are basis
vectors, rather than unit vectors, so that their lengths are not, in
general, equal to unity (see Weinreich 1998).  The corresponding unit vectors can trivially
be written as 
\be
\hat n_j = h_j \underline{\epsilon}_j\,,
\ee
where the corresponding scale factors 
$h_j = 1/|\underline{\epsilon}_j|$ are given by 
\be
h_p = { \xi \,(\xi - 1) \over \sqrt{A^2+(\xi-1)^2}}\,,
\ee
\be
h_q = { \xi \over \sqrt{A^2+(\xi-1)^2}}\,,\ee
and
\be
h_\theta = {\xi}\,.
\ee
In turn,
\be
\hat n_p = {A\over\sqrt{A^2+(\xi-1)^2}} \hat\xi + {\xi -1 \over\sqrt{A^2+(\xi-1)^2}} \hat\phi\,,
\ee
\be
\hat n_q = {\xi -1 \over\sqrt{A^2+(\xi-1)^2}} \hat\xi - {A\over\sqrt{A^2+(\xi-1)^2}} \hat\phi\,,
\ee
and
\be
\hat n_\theta = \hat\theta\,.
\ee

Parametric equations for magnetic field lines (lines of constant $q$) 
are easily obtained, and take the form
\be
x_B(\xi) = \xi \cos \left[{\xi - 1 - \ln  \xi - q \over A \sin\theta_0}\right]\,
\sin\theta_0\,,
\ee
\be
y_B(\xi) = \xi \sin \left[{\xi - 1 - \ln \xi  - q \over A\sin\theta_0}\right]\,
\sin\theta_0\,,
\ee
\be
z_B(\xi) = \xi \,\cos \theta_0\,,
\ee
where $q = -A \phi_0 \sin \theta_0$ sets the location ($1,\phi_0,\theta_0$)
where the corresponding  field line intersects the stellar surface.   Similar expressions
can also be found to parameterize lines of constant $p$ (though as noted
above, these lines cannot be traced to the surface). 
Figure 1 shows two field lines for $A = 2$ (dotted curves)  and $A = 20$ (solid curves) 
on the equatorial plane ($\theta_0 = \pi/2$) and a cone with $\theta_0 = \pi /4$.
As one can see,
the value of $A$ adjusts how tightly wound the field is, as can be characterized 
by a winding gap
\be
\Delta\xi_{w} \approx 2\pi A \sin\theta_0\,,
\ee
which represents the radial displacement between successive windings of the field.
\begin{figure}
\figurenum{1}
{\centerline{\epsscale{0.90} \plotone{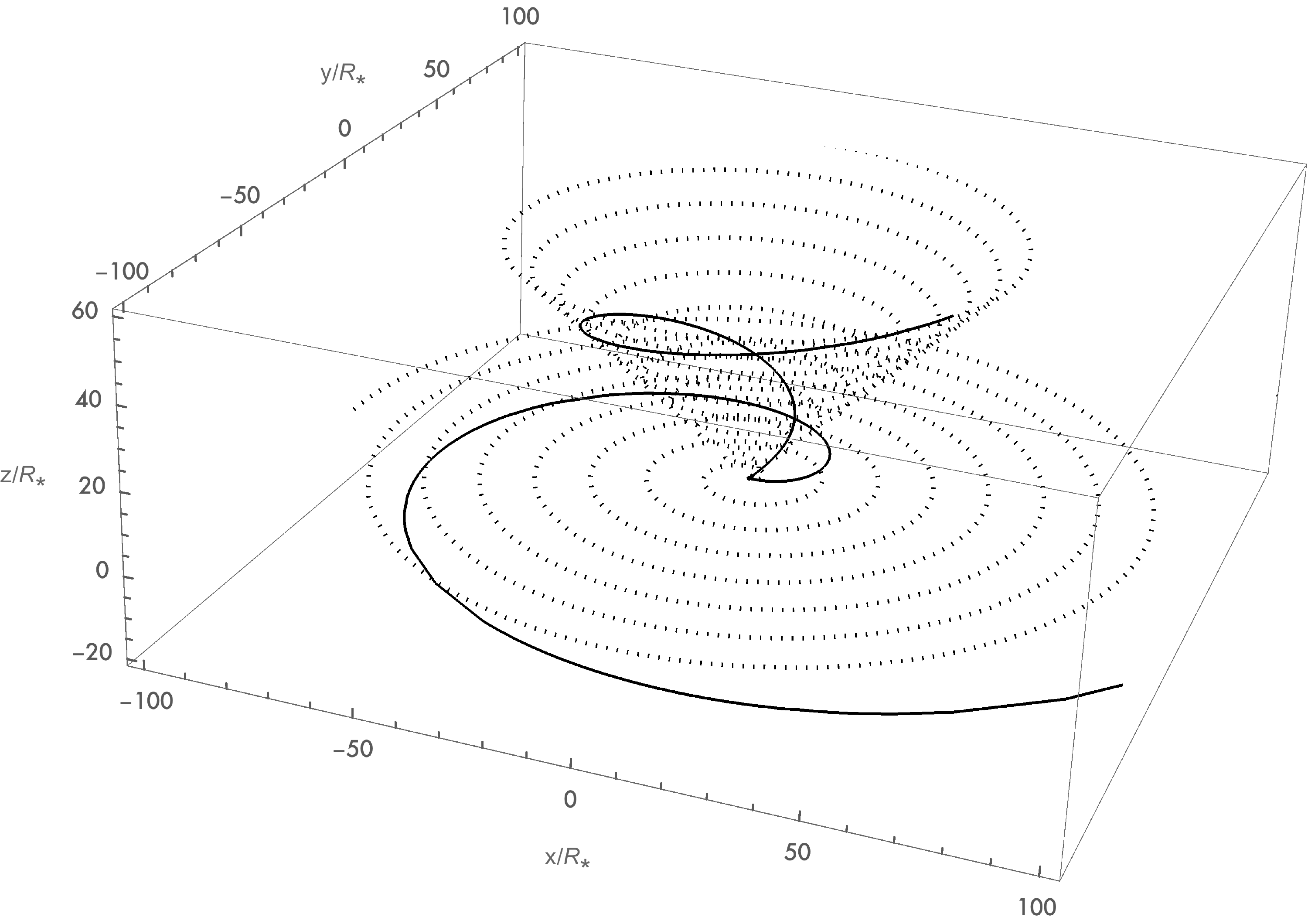} }}
\figcaption{Magnetic field lines for $A = 2$ (dotted curves) and
$A = 20$ (solid curves) for $\theta_0 = \pi/4$ (conical surface) and
$\theta_0 = \pi/2$ (equatorial plane).    }
\end{figure} 
Figure 2 shows two field lines for $A = 20$ (solid curves) for
a value of $\theta_0 = \pi /4$, and three lines of constant $p$ (dashed curves) on the
same  $\theta_0 = \pi /4$ cone.  
\begin{figure}
\figurenum{2}
{\centerline{\epsscale{0.90} \plotone{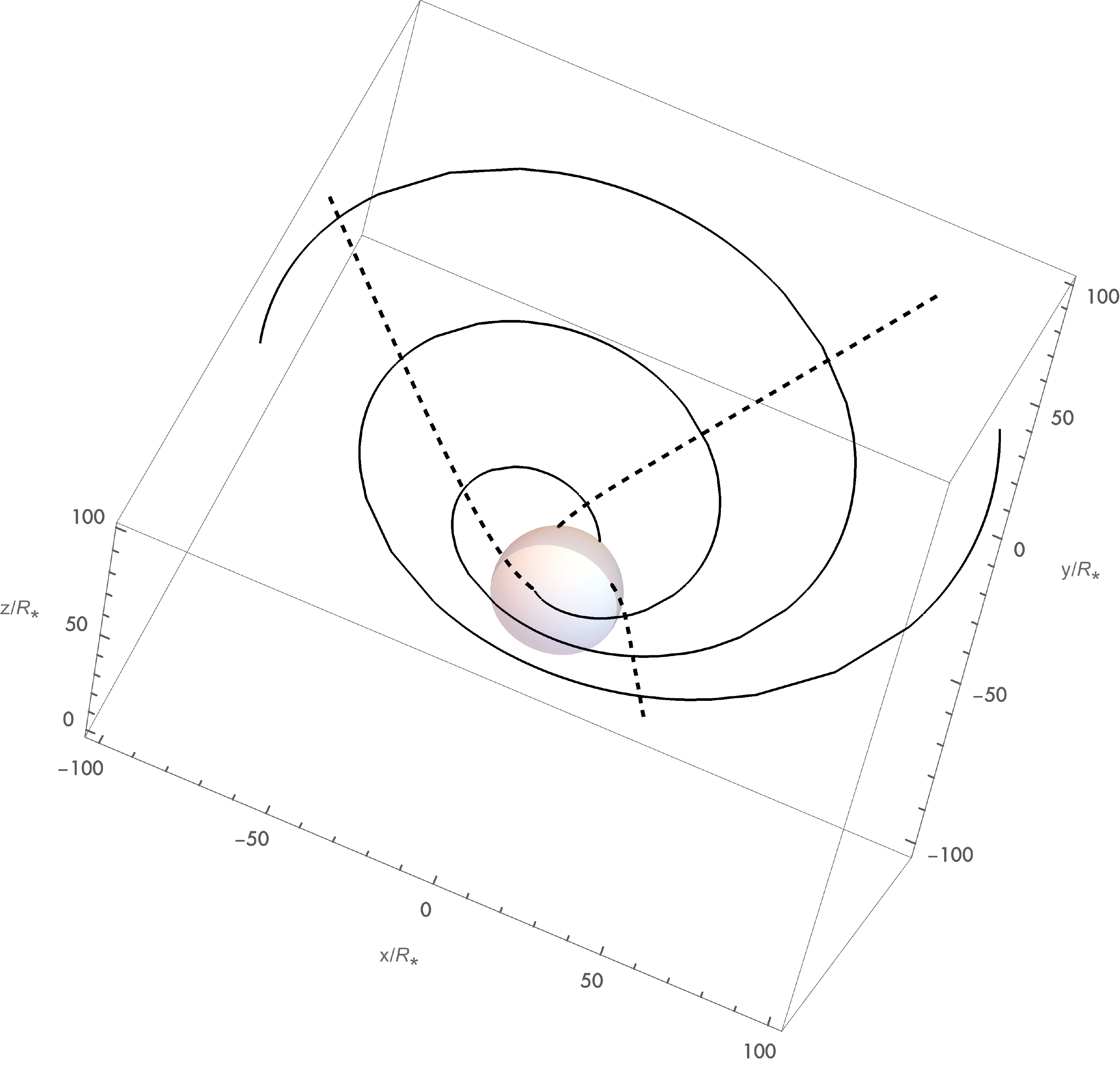} }}
\figcaption{Magnetic field lines  (solid curves) and lines of constant $p$  (dashed curves)
for $A = 20$ and $\theta_0 = \pi/4$ (conical surface).  The sphere at the center provides 
a point of reference and has radius $\xi = A = 20.$ }
\end{figure} 

\section{Magnetic Turbulence Structure}

Guided by previous work \citep{giajok94,casse,sullivan,fat10,fatadams},
we develop a formalism for modeling Alfv\'enic MHD turbulence by using
the orthogonal coordinate system developed in Section 2.  At present,
a complete theory of MHD turbulence in the interstellar medium remains
elusive. Nevertheless, it is generally understood that turbulence is
driven from a cascade of longer wavelengths to shorter wavelengths as
a result of wave-wave interactions. For strong MHD turbulence in a
uniform medium, this cascade seemingly produces eddies on small
spatial scales that are elongated in the direction of the underlying
magnetic field, so that the components of the wave vector $k_\perp$
and $k_{||}$ are related by the expression $k_{||} \propto
k_\perp^{2/3}$ (Sridhar \& Goldreich 1994; Goldreich \& Sridhar 1995;
Cho \& Lazarian 2003).  It is beyond the scope of this paper to extend
these results for our non-uniform geometry. Since our aim here is to
determine the possible effects of turbulence on cosmic ray propagation
into a star/disk system, we will assume a reasonable form for the
turbulent magnetic field as guided by basic principles.

Following the standard numerical approach for analyzing the
fundamental physics of ionic motion in a turbulent magnetic field, we
treat the total magnetic field ${\bf B}$ as a spatially turbulent
component $\delta{\bf B}$ superimposed onto the static Parker spiral
background field ${\bf B}_P$ described in Section 2. The turbulent field
$\delta {\bf B}$ is generated by summing over a large number of
randomly polarized waves with effective wave vectors $k_n$
logarithmically spaced between $k_1 = k_{min}$ and $k_N = k_{max}$,
\be
\delta {\bf B} = \sum_{n=1}^N \Lambda_n(p,q)  \,  \sin\left(k_n\,p 
+ \beta_n\right) [ \cos\alpha_n \hat n_q + \sin\alpha_n
\hat n_\theta]\,,
\label{turbform} 
\ee
where the direction and phase of each term is set through a random
choice of $\alpha_n$ and $\beta_n$ (e.g., Giacalone \& Jokipii 1994;
Casse et al. 2002; O'Sullivan et al. 2009; Fatuzzo et al. 2010), 
and $\Lambda_n(p,q)$ is chosen to maintain the divergenceless condition
(e.g., Equations 27 -- 29) 
and to set the turbulence profile (e.g. Equation 30).  Each
term in the sum represents an Alfv\'enic wave in the sense that
$\delta {\bf B}_n \perp {\bf B}_P$. Since the Alfv\'en speed $v_A$ is
much less than that of the relativistic cosmic rays, we can adopt a
static turbulent field (essentially a snapshot of the field
configuration) for calculating the effects on particle motion.  This
simplification then removes the necessity of specifying a dispersion
relation for each term.  Although the discrete formalism adopted here
represents an idealization of a continuous physical system, earlier
works indicate that 25 terms per decade in $k$ space yields a robust
description (e.g., Fatuzzo et al. 2016).  We therefore take the number 
of terms in the sum of equation (\ref{turbform}) to be 
$N = 25 \log_{10} (k_{max}/k_{min})$ throughout our work.

Since the variable $p$, which denotes the position along a field line,
is undefined at $\xi = 1$, an inner boundary for our analysis must be
set arbitrarily.  Toward that end, we combine equations (3) and (7)
for $q = 0$ to get the expression 
\be
p = A \ln\left(1-{1\over\xi}\right)+{\xi-1-\ln\xi \over A}\,,
\ee
which is then plotted in Figure 3 for $A = 2$, 10, and 40 (solid
curves), along with corresponding lines of the limiting form $p=\xi/A$
(dashed curves).  We note that for $\xi \ga A$, corresponding solid
and dashed curves become nearly parallel, so the above expressions for
$p$ are equivalent in our formalism.  We therefore set the inner
turbulence boundary for our system at $\xi_t = A$.

\begin{figure}
\figurenum{3}
{\centerline{\epsscale{0.90} \plotone{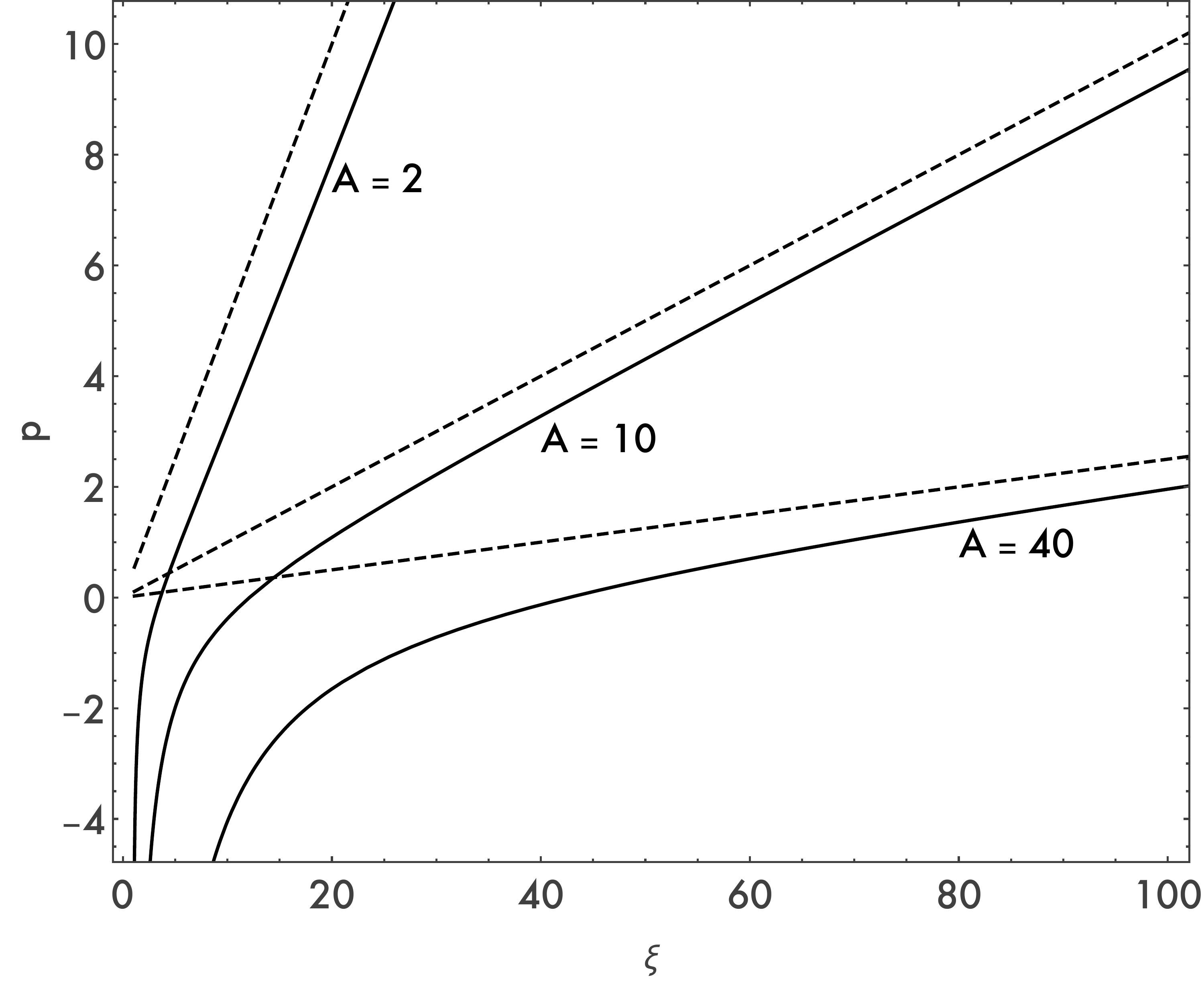} }}
\figcaption{The solid curves denote the value of $p$ as a function of
  $\xi$ for a representative equatorial field line on which $q = 0$
  for $A$ = 2, 10, and 40.  The dashed curves denote the corresponding
  values for the simple expression $p = \xi/A$.  }
\end{figure} 

To connect the wavevectors $k_n$ to their corresponding wavelengths
$\lambda_n$, we first note that in the limit $\xi \ga A$, $\Delta p
\approx \Delta \phi \sin\theta_0$, which means that each term in the
sum given by equation (20) represents a wave with $k_n \sin\theta_0$
wavelengths per winding of the background field. On the equatorial
plane, this result can be couched in terms of the path length $s$ (in
units of $R_*$) along a field line found by integrating the quantity  
\be
ds  =  \sqrt{d\xi^2 + \xi^2 d\phi^2} = \sqrt{{A^2+(\xi-1)^2 \over A^2}} d\xi\,,
\ee
where the condition that $q$ remains constant on a field line has been
used to relate $d\xi$ and $d\phi$.  Integrating from the turbulence
boundary (so that $s = p = 0$ at $\xi = \xi_t = A$) then yields the
expression 
\be
s  = { (\xi - 1)\sqrt{A^2+(\xi-1)^2} -  (A - 1)\sqrt{A^2+(A-1)^2} \over 2A}  
+ {A \over 2}  
\ln \left[{\xi-1+\sqrt{A^2+(\xi-1)^2} \over A-1+\sqrt{A^2+(A-1)^2}}\right] \,.
\ee
Coupled with equation (21), one then finds the approximate expression 
for the path length 
\be
s \approx {\xi^2 - A^2 \over 2 A} \approx {A p^2 \over 2}\,,
\ee
which for $A = 20$, yields a value of $p$ that is within $0.2$ of the
true value for all s.   

In Figure 4 we plot the sinusoidal dependence $\sin[k p]$ as a function of $s$ for an $A = 20$ 
equatorial field line for 
cases where $k = 1$ and $k = 3$.   The thin vertical lines denote locations where
the field has wrapped around by 1, 2, 3, and 4 times, respectively (e.g, where
$\phi$ has advanced by 2$\pi$, 4$\pi$, 6$\pi$, and 8$\pi$).  As expected, one clearly sees that to a good approximation, 
there is one complete wavelength
per winding for the $k = 1$ mode, and three complete wavelengths per winding for the $k = 3$ mode.
\begin{figure}
\figurenum{4}
{\centerline{\epsscale{0.90} \plotone{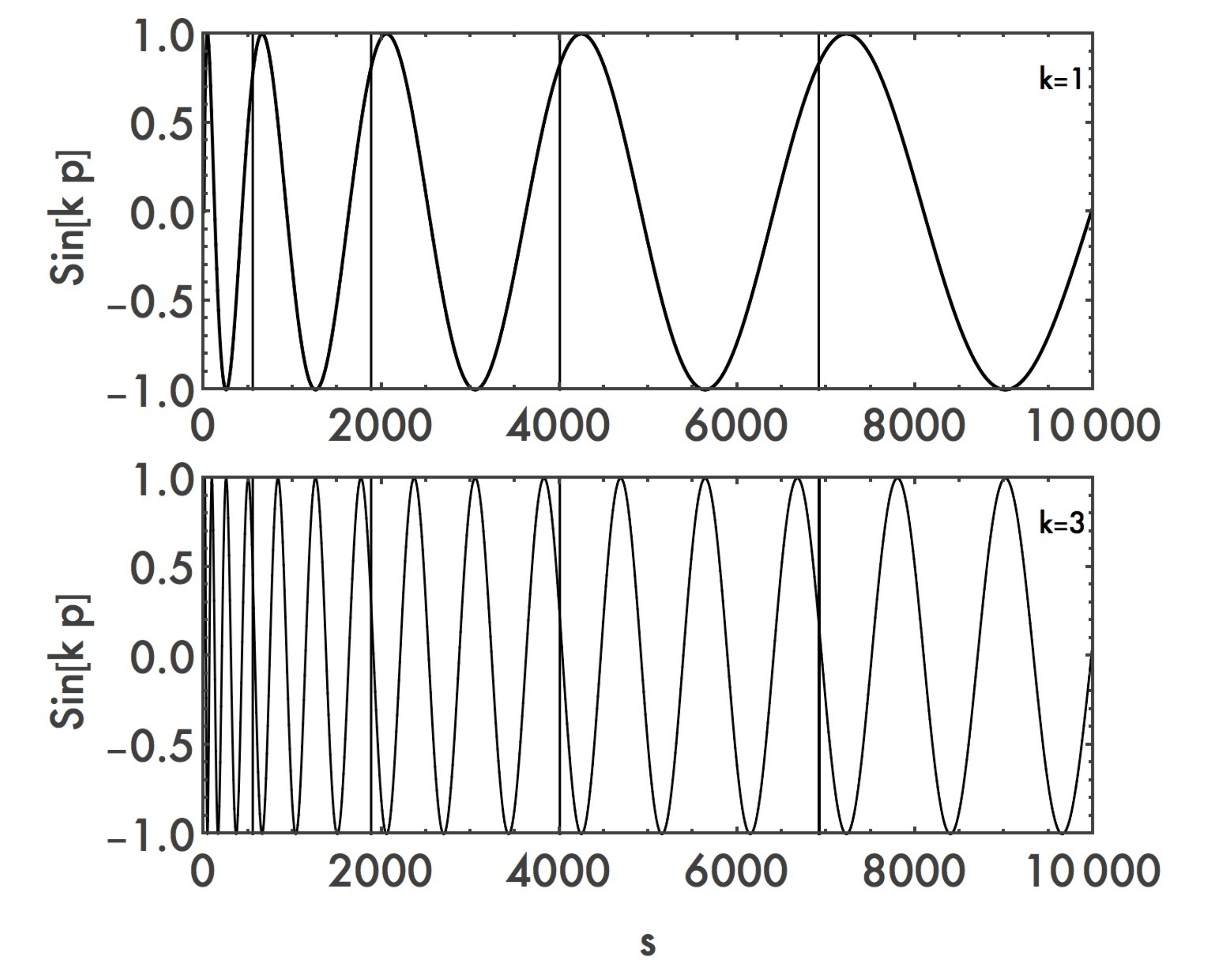} }}
\figcaption{ The function $\sin [k p]$ as a function of path length $s$ along a field line on the equatorial plane
with $A = 20$, for $k = 1$ (upper panel) and $k = 3$ (lower panel).   The thin vertical lines 
at $s = 548, 1879, 4001$, and $6913$ denote locations where
the field line has wrapped around by 1, 2, 3, and 4 times, respectively (e.g, where
$\phi$ has advanced by 2$\pi$, 4$\pi$, 6$\pi$, and 8$\pi$).
 }
\end{figure} 

Although the wavelength of a given mode (as specified by $k_n$)
increases with $s$, it is still useful to define an ``effective
wavelength" $\lambda_{n} (\xi)$ as equal to the displacement $\Delta
s$ along an equatorial field line corresponding to $k_n \Delta p =
2\pi$ (hence, $\lambda_n$ is also in units of $R_*$).  Using equation
(24), one finds 
\be
\lambda_n \equiv \Delta s \approx {A\over 2} \left[2 p \Delta p + (\Delta p)^2\right]
 \approx \left[ {2\pi \over k_n} \sqrt{\xi^2 -A^2} + \lambda_{t;n} \right]\,,
\ee
where the effective wavelength at the turbulence boundary is given by 
\be
\lambda_{t;n} = {2 \pi^2 A \over k_n^2}\,.
\ee

To complete the analysis, we ensure that the no-monopole condition
$\nabla\cdot \delta {\bf B}_n = 0$ is satisfied by selecting an
appropriate form of the coefficients $\Lambda_n$.  Toward that end, we
use the divergence operator in our $(p,q,\theta)$ coordinate system to
write the no-monopole condition as 
\begin{eqnarray}
\nabla\cdot \delta {\bf B}_n 
= {\sin(k_n p+\beta_n) \cos(\alpha_n) \over h_p h_q h_\theta} 
\left[{\partial\over\partial q}  (h_p h_\theta \Lambda_n)\right]  = 0\,,
\end{eqnarray}
which requires
\be
h_p h_\theta \Lambda_n = f(p)\;.
\ee
Using equations (10) and (12), and
setting $f(p)$ equal to the constant 
$\Lambda_{t;n}$, one then finds 
\be
\Lambda_n(\xi) = \Lambda_{t;n} {A^2 (A-1)\over \xi^2 \, (\xi-1)}\ 
{\sqrt{A^2+(\xi-1)^2} \over\sqrt{A^2+(A-1)^2} }\;,
\ee
where the expression has been normalized so that $\Lambda_n(A) =
\Lambda_{t;n}$.  We note that $\Lambda_n \rightarrow \xi^{-2}$ when
$\xi \gg A$, and in turn, $\delta B/ B_P \rightarrow \xi^{-1}$.

The desired spectrum of the turbulent magnetic field is set
through the appropriate choice of scaling for an assumed turbulent 
profile, i.e., 
\begin{equation}
\Lambda_{t;n}^2 = \Lambda_{t;1}^2\left[{k_n \over k_1} \right]^{-\Gamma}
{\Delta k_n\over \Delta k_1} 
= \Lambda_{t;1}^2\left[{k_n \over k_1} \right]^{-\Gamma+1}\,,
\end{equation}
where $\Gamma = 5/3$ is used for
Kolmogorov turbulence.  We note that for our logarithmic binning
scheme, the value of $\Delta k_n/k_n$ is the same for all values of
$n$.  The value of $A_{t;1}$ is set by an amplitude parameter $\eta$ 
that specifies the average energy density of the turbulent field with
respect to the background Parker-spiral field at the inner turbulence boundary; 
specifically, $\eta$ is defined through the expression  
\be
\eta = {\langle \delta B^2\rangle \over B_{p;t}^2}\,,
\ee
where $B_{P;t}$ is the magnitude of the equatorial Parker spiral magnetic field at $\xi = A$
(our inner boundary for turbulence).
In order to determine the entire structure of the 
magnetic field, we must specify the parameters $B_*$ (which then sets $B_{P;t}$), $A$,
$k_1$, $k_N$, $\eta$, and $\Gamma$.  We note, however, that 
since charged particle motion through a turbulent field is a resonant phenomena, the
dynamics are not sensitive to the choice of $k_N$, so long as the particle
radius of gyration falls within the range of largest and smallest fluctuations.  
Turbulent field lines are illustrated in Figure 5 for the case $A = 20$,
$k_1 = 1$, $k_N = 10^4$ (so that $N = 100$), $\eta = 1$, and $\Gamma = 5/3$.
\begin{figure}
\figurenum{5}
{\centerline{\epsscale{0.90} \plotone{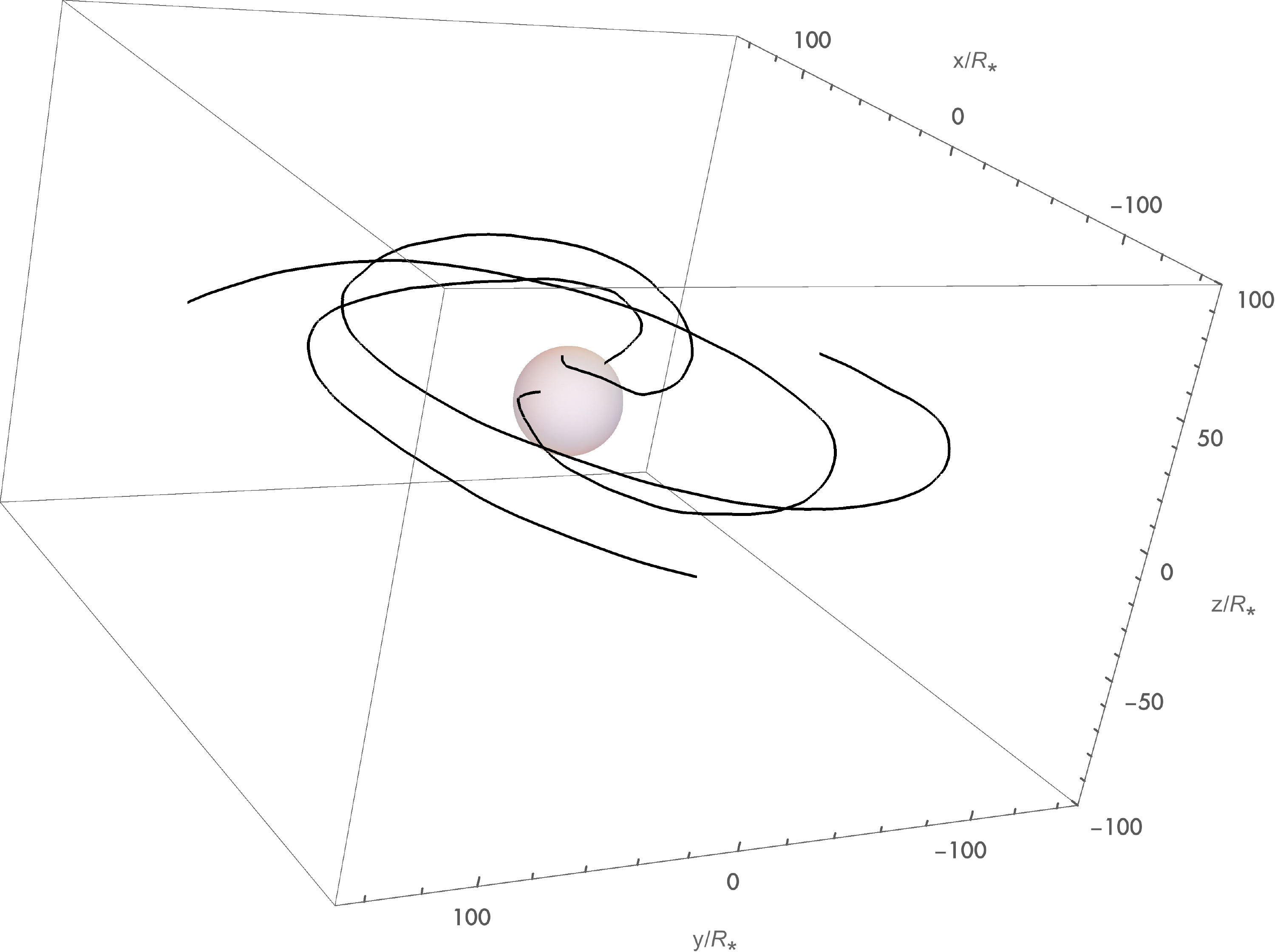} }}
\figcaption{Turbulent field lines for the case $A = 20$,
$k_1 = 1$, $k_N = 10^4$, $\eta = 1$, and $\Gamma = 5/3$. 
The sphere at the center represents the turbulence boundary,
and has radius $\xi = A = 20.$  
 }
\end{figure} 

\section{Particle Dynamics}

The aim of this paper is to determine how cosmic-rays propagate
through the magnetized disk environments surrounding T Tauri stars. We
limit our analysis to relativistic protons, but note that our results
are easily scaled to other cosmic-ray species.  The general equations
that govern the motion of protons with Lorentz factor $\gamma$ through
a magnetic field take the well-know form 
\be
{d {\bf u}\over  dt} =  {c\over R_0}
\,{{\bf u} \times {\bf b} \over \gamma}
\qquad {\rm and} \qquad 
{d  {\bf  r}\over dt} = c {{\bf u}\over \gamma}\;,
\ee
where ${\bf u} = \gamma {\bf v}/c$, ${\bf b} = {\bf B} /B_*$, and
\be
R_0 \equiv {m_p c^2 \over e B_*} = 3.1\times 10^{3} \,{\rm cm}
\left({B_* \over 10^3 \,\hbox{G}}\right)^{-1}\,.
\ee
The kinematics equations are readily solved given specified
initial conditions (${\bf u_i} $ and ${\bf r_i}$) for our magnetic field geometry
using standard numerical methods.  
For completeness, 
we note that the radius of gyration for a relativistic particle ($v \approx c$) moving through the spiral field
at location $\xi$ with a pitch angle $\alpha_p$ is 
\be
 R_g(\xi) ={\gamma R_0 B_* \sin \alpha_p \over B(\xi)} \,.
\ee
Cosmic-rays can only be funneled toward the T Tauri star if $R_g(\xi)/R_* \ll \xi$, which requires
\be
\gamma \ll 4.8 \times 10^7 {\sqrt{A^2 +(\xi-1)^2} \over A \xi} \left({B_* \over 10^3\, {\rm G}}\right)
\,\left({R_* \over 1.5\times 10^{11} \,{\rm cm}}\right)\,
\ee

Important
insight can be gleaned by considering the motion of particles in 
the limit $\xi \gg A$, for which ${\bf b}\approx (A\xi)^{-1} \hat \phi$.  
In this limit, the magnetic field has the same form as that of an infinite, straight, current-carrying wire.
As shown in Aguierre et al. (2010), charged particles moving through such a field that have a radius
of gyration smaller than their distance to the central wire will follow the field lines around the
wire, but also exhibit a drift in the direction of the current.  We extend the analysis for relativistic particles in Appendix B, and reach a similar conclusion.  We do find,
however, that the drift speed is proportional to the Lorentz factor $\gamma$ (see equation [B10]).
These results are further confirmed numerically, as illustrated in Figure 6.  In our analysis, we will
therefore consider particles with a small enough Lorentz factor that their drift does not carry them 
a distance $R_*$ farther than the equatorial plane.   Note that most of the ionization is expected to result from 
mildly relativistic particles: The cosmic ray flux increases 
rapidly with decreasing energy down to $E\sim1$ GeV 
(\citealt{webber,moskalenko}), and ionization is efficient for 
this energy range (see Umebayashi \& Nakano 1981, 
Padovani et al. 2009, 2011; Cleeves et al. 2013). 
As a result, our results are expected to be robust.

\begin{figure}
\figurenum{6}
{\centerline{\epsscale{0.90} \plotone{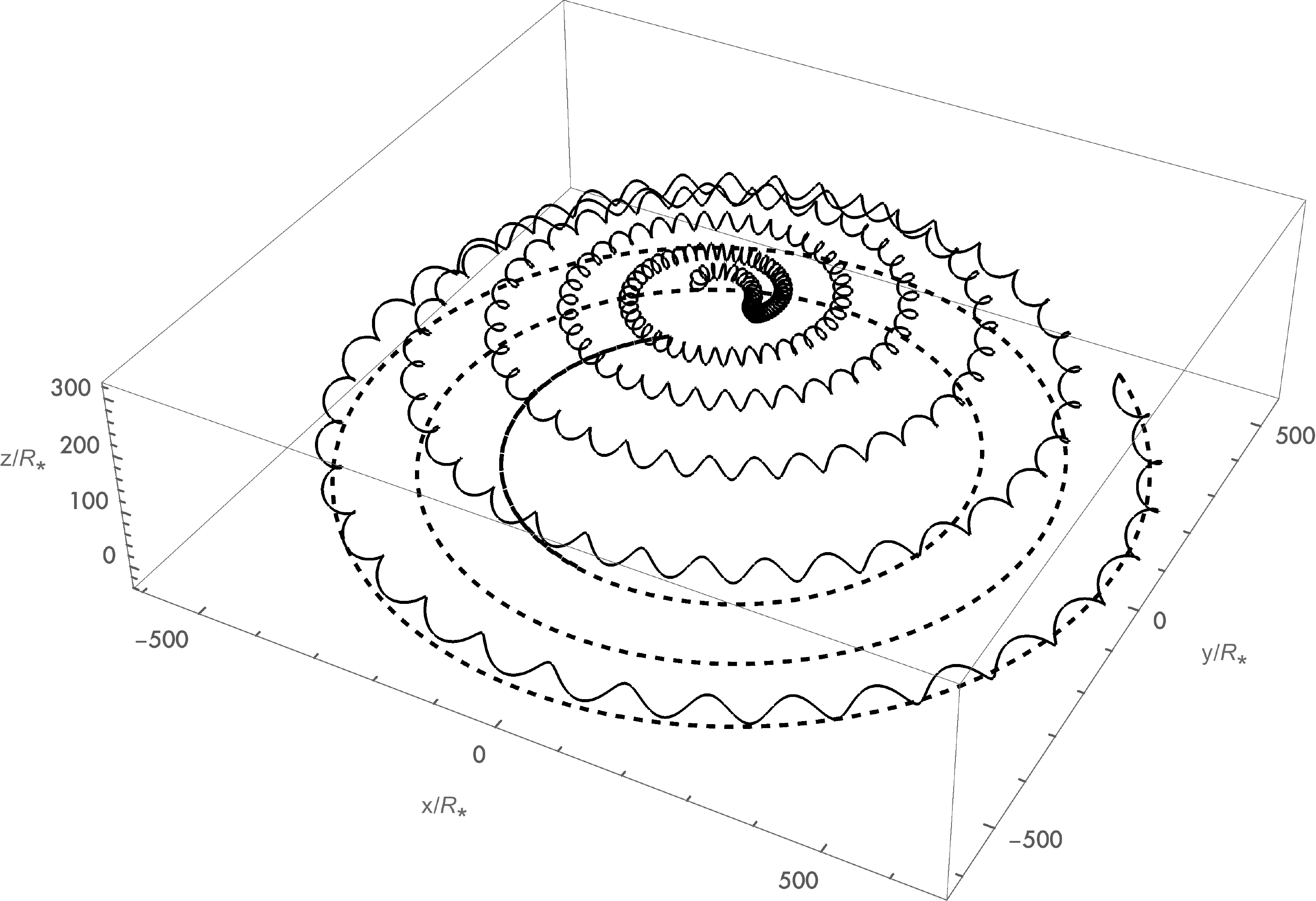} }}
\figcaption{Trajectories for particles with
Lorentz factors  $\gamma = 10^5$ (solid curve) and $\gamma = 10^3$
(dashed curve) launched from the equatorial plane of a Parker spiral magnetic field
with $A = 20$ and adopted fiducial values of  $B_* = 10^3$ G
and $R_* = 1.5 \times 10^{11}$ cm.
 }
\end{figure} 

\section{Effect of turbulence on cosmic-ray reflection}

The magnetic moment of a relativistic proton
\be
\mu = {\gamma^2 m_p\,  v^2 \sin^2\alpha_p\over 2B}\,,
\ee
is an adiabatic invariant under the condition that the field does
not change significantly within a gyration radius, i.e., in the limit 
\be
{\gamma}  \ll {1 \over R_0 B_*} {B^2\over |\nabla B|} = 4.8 \times 10^7{ \left[A^2+(\xi-1)^2\right]^{3/2} \over A \xi (2+2A^2-3\xi +\xi^2)} \left({B_* \over 10^3\, {\rm G}}\right)
\,\left({R_* \over 1.5 \times 10^{11} \,{\rm cm}}\right)\, \,,
\ee 
which similar to the condition expressed by equation (35).
However, as is clear from Figure 6, a much stricter condition in our analysis is placed on the particle
Lorentz factor by requiring that the particle drift is negligible in our numerical experiments. 

Since the Lorentz factor of a particle remains constant in a 
time-independent magnetic field, the adiabatic invariance can be
expressed as
\be
{\sin^2 \alpha_p\over B} = {\rm constant}\, .
\ee 
As a charged cosmic-ray moves toward the
star, its pitch angle must increase to match the increasing
field strength.  But since  $\sin\alpha_p \le 1$, the cosmic-ray must eventually be reflected 
at a mirror point where $\alpha_p = \pi/2$.  To reach the turbulence boundary, 
a proton must therefore be injected from a radius $\xi_i$ with a pitch
angle smaller than 
\be
\alpha_{crit} = \sin^{-1} \left[{A \left[A^2+ (\xi_i-1)^2 \right]^{1/4} \over \xi_i \left[A^2+ (A-1)^2 \right]^{1/4} }\right]\,.
\ee
For injected pitch angles $\alpha_i > \alpha_{crit}$, reflection
occurs at a radius $\xi_r$ that can be well-approximated by solving
the equation 
\be
\xi_r^4 (A^2+\xi_{i}^2) - \xi_r^2 \xi_{i}^4 \sin^4\alpha_i - A^2\xi_i^4 \sin^4\alpha_i = 0\,.
\ee
Figure 7 illustrates the location where reflection occurs as a
function of injection pitch angle for particles injected from a radius
$\xi_i = 10^4$ on the equatorial plane of a (turbulence-free) Parker
spiral field characterized by $A = 2$ (solid curve), $A = 20$ (dashed
curve) and $A = 40$ (dotted curve).

\begin{figure}
\figurenum{7}
{\centerline{\epsscale{0.90} \plotone{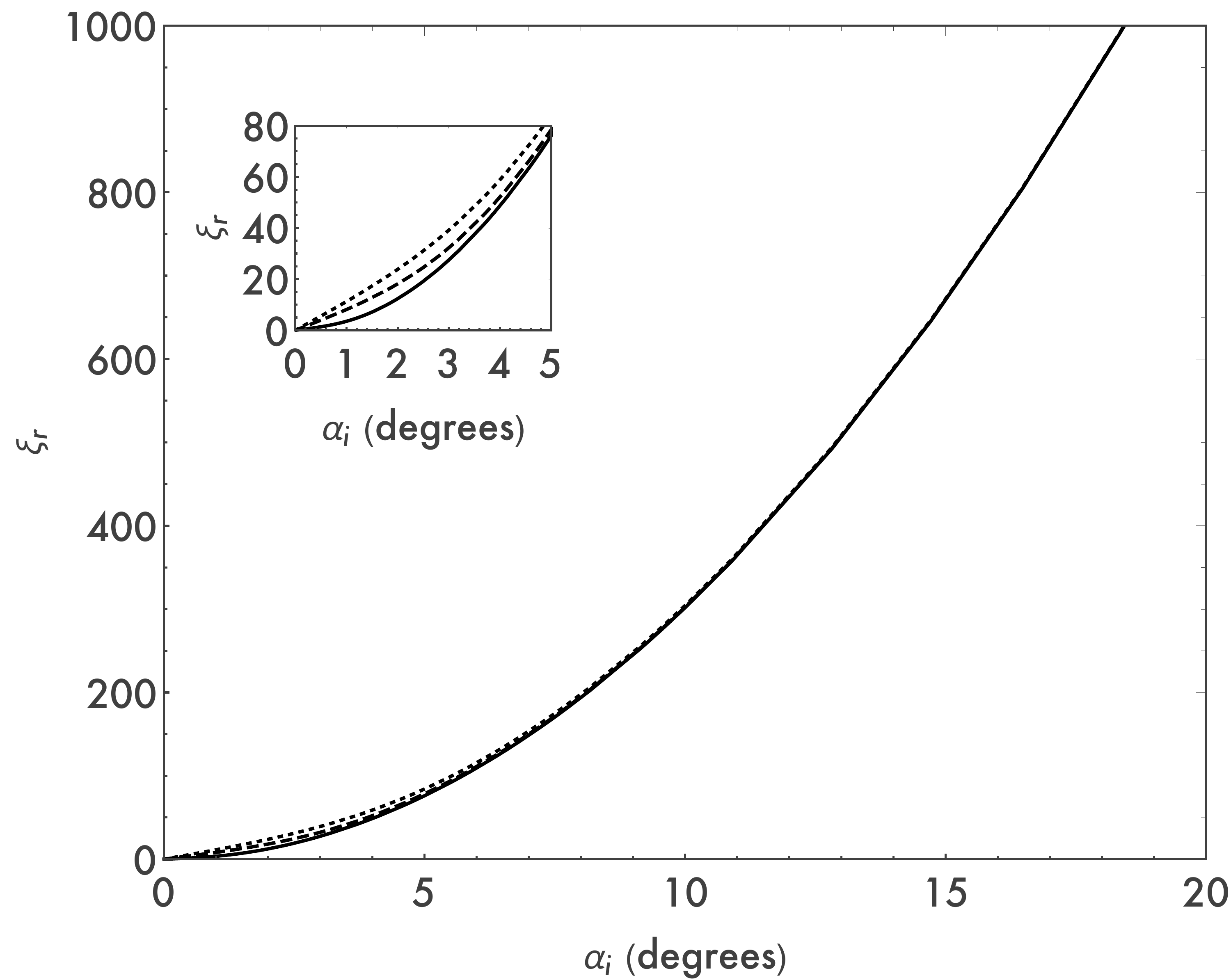} }}
\figcaption{Reflection radius for $A = 2$ (solid curve), $A = 20$
  (dashed curve), and $A = 40$ (dotted curve) for a particle injected
  with pitch angle $\alpha_i$ at a radius $\xi_i = 10^4$.  }
\end{figure} 

We investigate the effect of turbulence on reflection by running a
suite of numerical experiments under different magnetic field scenarios.  
Each experiment is defined in terms of a specific magnetic profile (as
set by the choice of $A$, $k_1$, $\Gamma$ and $\eta$), particle
energetics (as defined by $\gamma$), and an injection scenario.  The
value of $k_N$ is set to ensure that the minimum wavelength is an
order of magnitude smaller than the particle radius of gyration, both
of which scale as $\xi$ (see discussion at the end of Section 3).  
We note from equation (25) that $\lambda_{min} \la 2\pi\xi / k_N$ when
$\xi \ga A$, and from equations (6) and (34) that $R_g \approx \gamma
A R_0 \xi \sin\alpha$.  We therefore set 
\be
k_{N} = 10 \left[  {2\pi \xi R_* \over R_g}\right] = 
{3\times 10^9 \over \gamma A \sin\alpha_{crit}}\,
\left({R_*\over 1.5\times 10^{11}\,{\rm cm}} \right)\,\left({B_*\over 10^3\,{\rm G}} \right)\,,
\ee
thereby invoking the smallest possible particle pitch angle.  The
value of $k_1$ is constrained by requiring that the longest effective
boundary wavelength $\lambda_{t;1}$ be smaller than the radial extent
of the T Tauriosphere, where, like our Heliosphere, the pressure from
the stellar wind is balanced by ISM pressure.  This boundary occurs at
a radius that is on the order of 1000 AU (compared to $\sim 120$ AU
for the Heliopause).  Guided by equation (26), we thus invoke the constraint 
\be
k_1 \ge  k_{crit} \equiv\sqrt{{ 2\pi^2 A \over 10^5} 
\left({R_* \over 1.5\times 10^{11} \,{\rm cm}}\right)}\,.
\ee
We adopt $k_1 = 10 k_{crit}$ for most of our cases, but also consider
larger and smaller values.

In addition to specifying the field structure, one must also specify
the distribution of injected particles being considered.  As noted
above, most of the ionizing potential of cosmic-rays comes from
moderately relativistic particles.  However, the computational time
required to complete the numerical experiments in our analysis for
$\gamma\approx 1$ particles makes this choice of particle energies
impractical.  Luckily, the reflection point is not sensitive to the
particle energy so long as particle drifting remains small, which then
ensures that the conditions expressed in equations (35) and (37) are
met.  Taking everything into account, we adopt a Lorentz factor of
$\gamma = 10^2$.

Ideally, it makes sense to inject particles at the outer edge of the
T-Taurisphere, which as noted above, occurs at a radius $\sim 1000$
AU.  But given our choice of particle energy, the computational time
required to integrate inward from the T-Taurisphere once again makes
such a choice unfeasible.  Since $\delta B \sim \xi^{-1}$, the effects
of turbulence become increasingly significant as the particles gets
closer to the star.  We therefore arbitrarily set the injection
boundary at 100 AU (which for $R_* = 1.5\times 10^{11}$ cm,
corresponds to $\xi_i = 10^4$). As shown below, moving the injection
boundary from 1,000 AU to 100 AU does not significantly affect the
results, though care must be taken to properly interpret the injection
distributions being considered.

The experiments performed and their corresponding output measures are
listed in Table 1.  Each experiment consists of numerically
integrating the equations of motion for $N_p=300$ particles for a
specified field structure (as defined by $A$, $\eta$, $k_1$,
$\Gamma$), stellar parameters $R_*$ and $B_*$, and an injection
scenario, but with each particle experiencing a different realization
of the specified turbulent field as set through the random assignment
of $\alpha_n$ and $\beta_n$ in equation (20).  The mean and median of
the ensuing distribution of reflection radii are then used as the
output measures.  Given the large number of system parameters, we
consider only Kolmogorov turbulence ($\Gamma = 5/3$), and adopt
fiducial values of $B_* = 10^3$ G and $R_* = 1.5 \times 10^{11}$ cm.
We then explore how the parameters $\eta$, $A$, and $k_{min}$ affect
particle reflection via experiments 1 - 7, for which all particles are
injected at the critical angle (e.g.  $\alpha_i = \alpha_{crit}$).
Histograms of the resulting distributions of reflection radii are
presented in Figure 8.  In each panel, the mean of each
distribution is represented by a vertical line (as defined in the
figure caption), and can be compared directly to the known reflection
radius in the absence of turbulence, which occurs at $\xi_r = A$.

Experiment 8 assess the overall effect that turbulence has on the
ionization of a disk surrounding a T Tauri star.  We assume the same
turbulence profile as for Experiment 1, adopt a disk radius $R_d = 50$
AU (e.g., $\xi_d = 5,000$), and randomly pick the injection angle by
selecting $\mu_i = \cos\alpha_i$ from a flat-top distribution ranging
between $\cos\alpha_R$ and $\cos\alpha_c$ (where $\alpha_R = 45^o$ is
the injection pitch angle for which $\xi_r = \xi_d$ in a non-turbulent
field).  The histogram of the resulting distribution of reflection
radii is presented in Figure 9, along with the corresponding
distribution function in the absence of turbulence 
\be
f[\xi_r] = \left[ {1\over \cos\theta_c - \cos\theta_R} \right] {d\mu \over d\xi_r} \,,
\ee
easily obtained from equation (40).  The mean of the distribution for
Exp. 8 (represented by the dashed vertical line in Figure 9) can be
compared to its corresponding turbulent-free value 
\be 
\langle \xi_r \rangle = \int_A^{\xi_d} \xi_r f[\xi_r] d\xi = 2650\,.  
\ee 
Likewise, the median of the distribution (which appears in Table 1)
can be compared to its corresponding turbulent-free value via the
expression 
\be 
\int_A^{\xi_{med}} f[\xi_r] d\xi = 0.5\,, 
\ee 
which is easily calculated to be $\xi_{med} = 2730$.  While the mean
and median values of Experiment 8 are quite similar to their
non-turbulent counterparts, it is easily seen that turbulence does
reduce the number of cosmic rays that reach the inner part ($\xi \la
10^3$) of the disk.  Of course, only a small fraction of cosmic-rays
that enter the T-Taurisphere at 1000 AU (which for our choice of $R_*$
corresponds to $\xi = 10^5$) will reach the disk.  Specifically, for
the physical conditions used in Experiment 8, particles would have to
have a pitch angle of $\alpha_T \le 12.9^o$ at the T-Taurisphere in
order to reach the disk in the absence of turbulence.  As such, the
particles represented in Figure 10 account for only $f =
1-\cos(12.9^o) = 0.025$ of all the particles that enter the
T-Taurisphere for a Parker spiral with $A = 20$.

\bigskip 
\begin{table}
\centering
\begin{tabular} {c cc c l c c c} 
\hline
\hline
\, & \, & {\bf Summary} & {\bf of} & {\bf Experiments} \\ 
\hline
\hline
Exp. &$\alpha_i$ & $\eta$ & $A$ &$k_1$ & $\langle \xi_r \rangle$  & $\xi_{med}$ \\ [0.5ex]
\hline
1 &$\alpha_c$& 1& 20&$10 k_{crit}$& 105 & 97.4\\
2 &$\alpha_c$&$0.1$&20&$10 k_{crit}$& 35.8& 33.1\\
3 &$\alpha_c$&$0.01$&20&$10 k_{crit}$& 28.8& 30.1 \\
4 &$\alpha_c$ & $1$&10&$10 k_{crit}$&41.4&39.2\\
5 &$\alpha_c$ & $1$&40&$10 k_{crit}$&210&198\\
6 &$\alpha_c$ & $1$&20&$k_{crit}$&67.3& 54.3\\
7 &$\alpha_c$ & $1$&20&$100 k_{crit}$&237&229\\
8 &flat top& $1$&20&$10 k_{crit}$&2640 &2610 \\
\hline
\end{tabular}
\caption{Parameters are listed for each set of numerical experiments,
  including the pitch angle $\alpha_i$, the turbuence amplitude $\eta$,
  the magnetic field geometry parameter $A$, and the wavenumber
  $k_1$. The last two columns give the resulting values of the mean
  and median turning points of the distribution. }
\label{table:quasar}
\end{table}

\begin{figure}
\figurenum{8}
{\centerline{\epsscale{0.90} \plotone{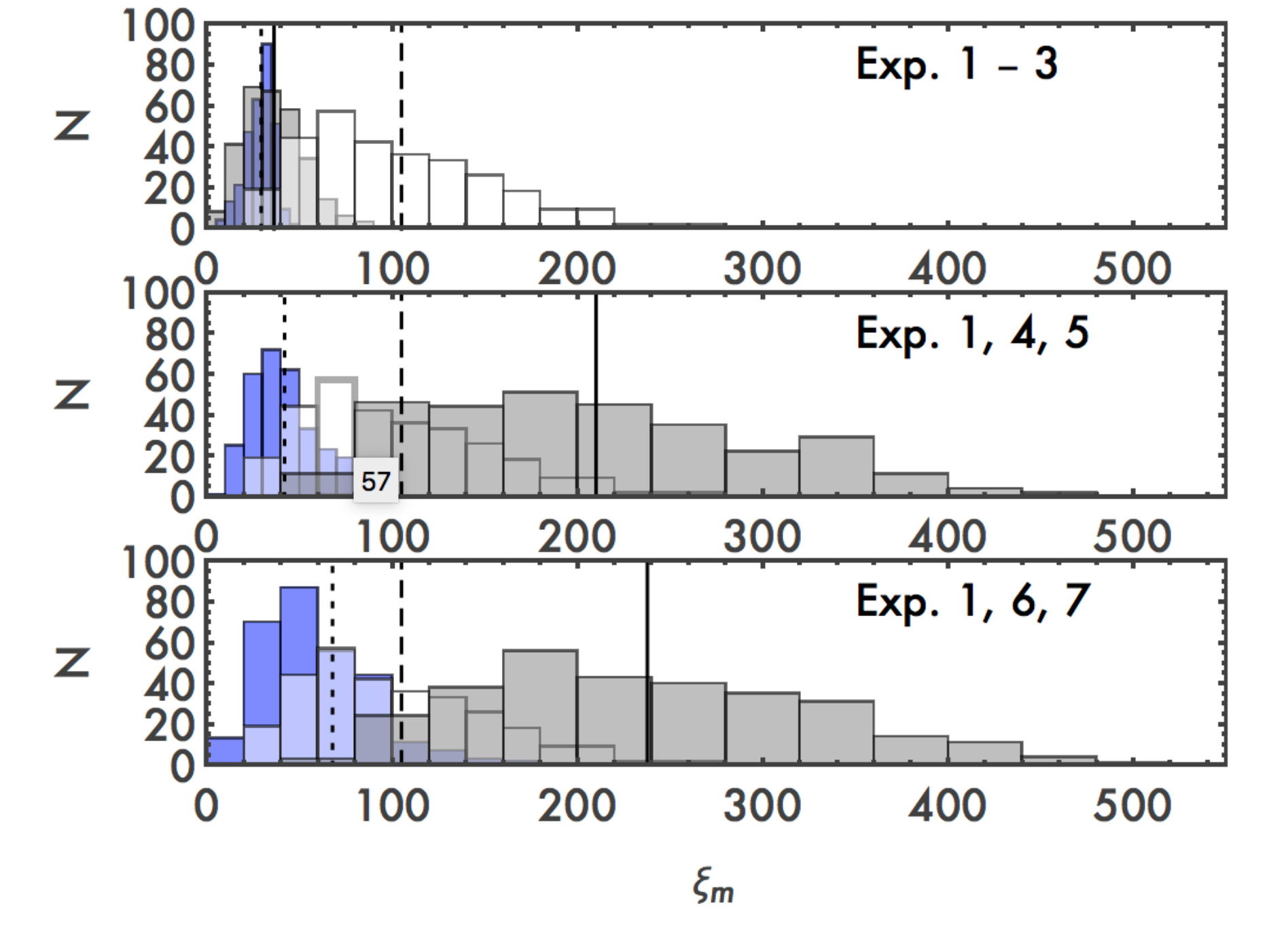} }}
\figcaption{Distributions and corresponding means of reflection radii.  Top panel - results for Exp. 1 (white histogram - dashed vertical line), 
Exp. 2 (gray histogram - solid vertical line) and Exp. 3 (blue histogram - dotted vertical line), for which
the turbulence strength varies respectively from $\eta = 1$, 0.1, and 0.01. Middle panel - results for Exp. 5 
(gray histogram - solid vertical line), Exp. 1 (white histogram - dashed vertical line) 
and Exp. 4 (blue histogram - dotted vertical line), for which
the magnetic profile parameter $A$ varies respectively from $A = 40$, 20, and 10.  Bottom panel - results for Exp. 7 
(gray histogram - solid vertical line), Exp. 1 (white histogram - dashed vertical line) 
and Exp. 6 (blue histogram - dotted vertical line), for which
the magnetic profile parameter $k_{min}$ varies respectively from $k_{min} = 100 k_{crit}$, $10 k_{crit}$, 
and $k_{crit}$. 
 }
\end{figure}

\begin{figure}
\figurenum{9}
{\centerline{\epsscale{0.90} \plotone{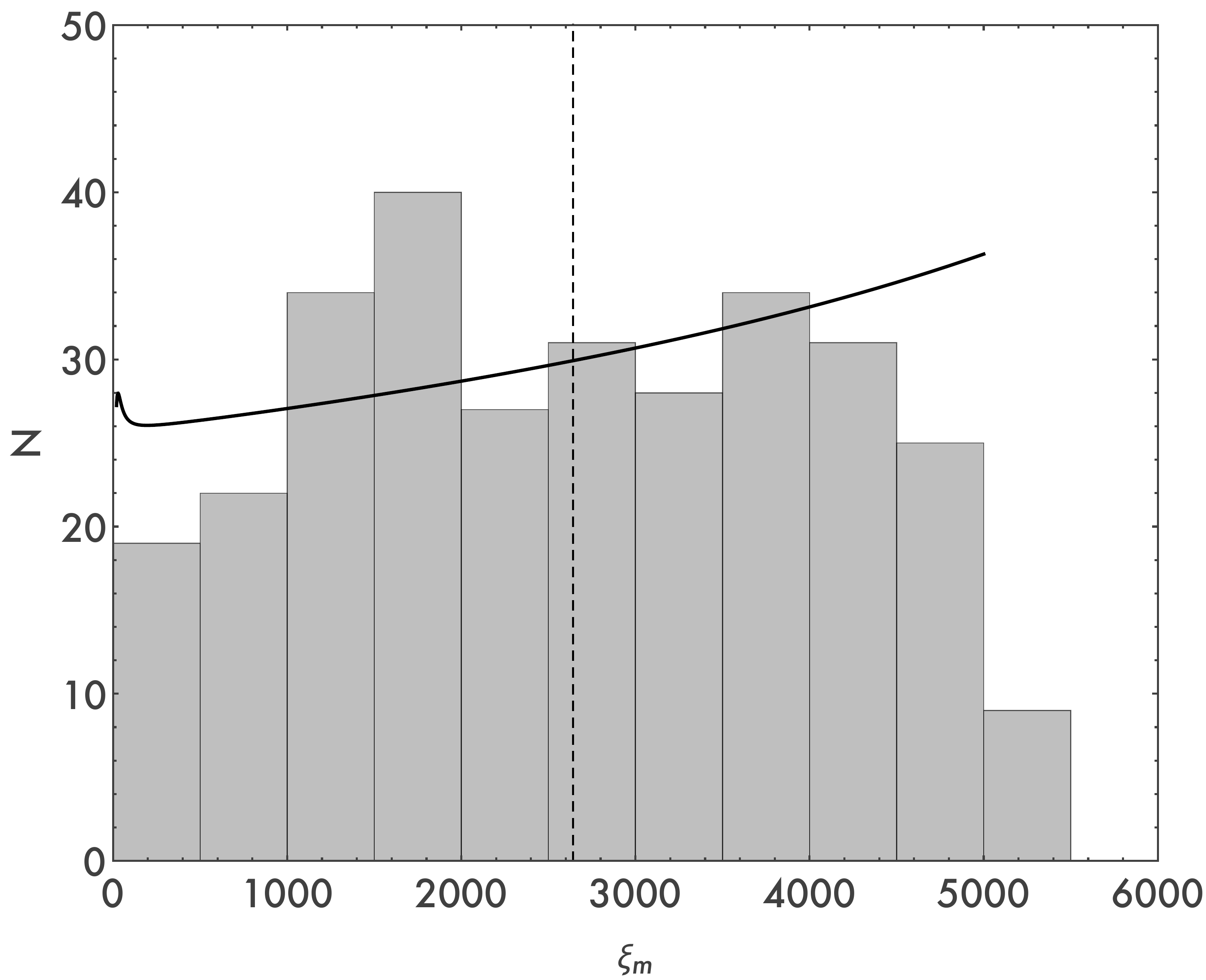} }}
\figcaption{Distributions of reflection radii for Exp. 8.  The vertical line denotes the mean of the distribution, 
and the solid curve represents the ``normalized" distribution function $f_r[\xi_r]$ for a non-turbulent field. 
 }
\end{figure}

To  assess the impact of our choice of injection radius,  we repeat Experiment 1, but inject
particles at at radius of $\xi_i = 2\times 10^4$ (200 AU).
The 
ensuing distribution of pitch angles (with respect to the turbulent field) is then 
obtained at $\xi =10^4$ (100 AU), with the results shown in Figure 10.  As expected, the presence of turbulence 
spreads out the distribution of pitch angles about the non-turbulence field value (depicted
by the vertical line).  Not surprisingly, this spread is quite narrow owing to the fact that
the turbulent field scales as $\delta B \sim \xi^{-1}$.  If these particles continued
spiraling inward in a non-turbulent field, the corresponding distribution of 
reflection radii $\xi_r$  would range between 17 and 25, 
which is much narrower than 
the distribution of reflection radii obtained in Experiment 1.  As such, the distributions
obtained in our experiments, for which particle injection occurred at $\xi_i = 10^4$,
would be very similar to those that would have been obtained had we injected the particles
at $\xi = 10^5$.
\begin{figure}
\figurenum{10}
{\centerline{\epsscale{0.90} \plotone{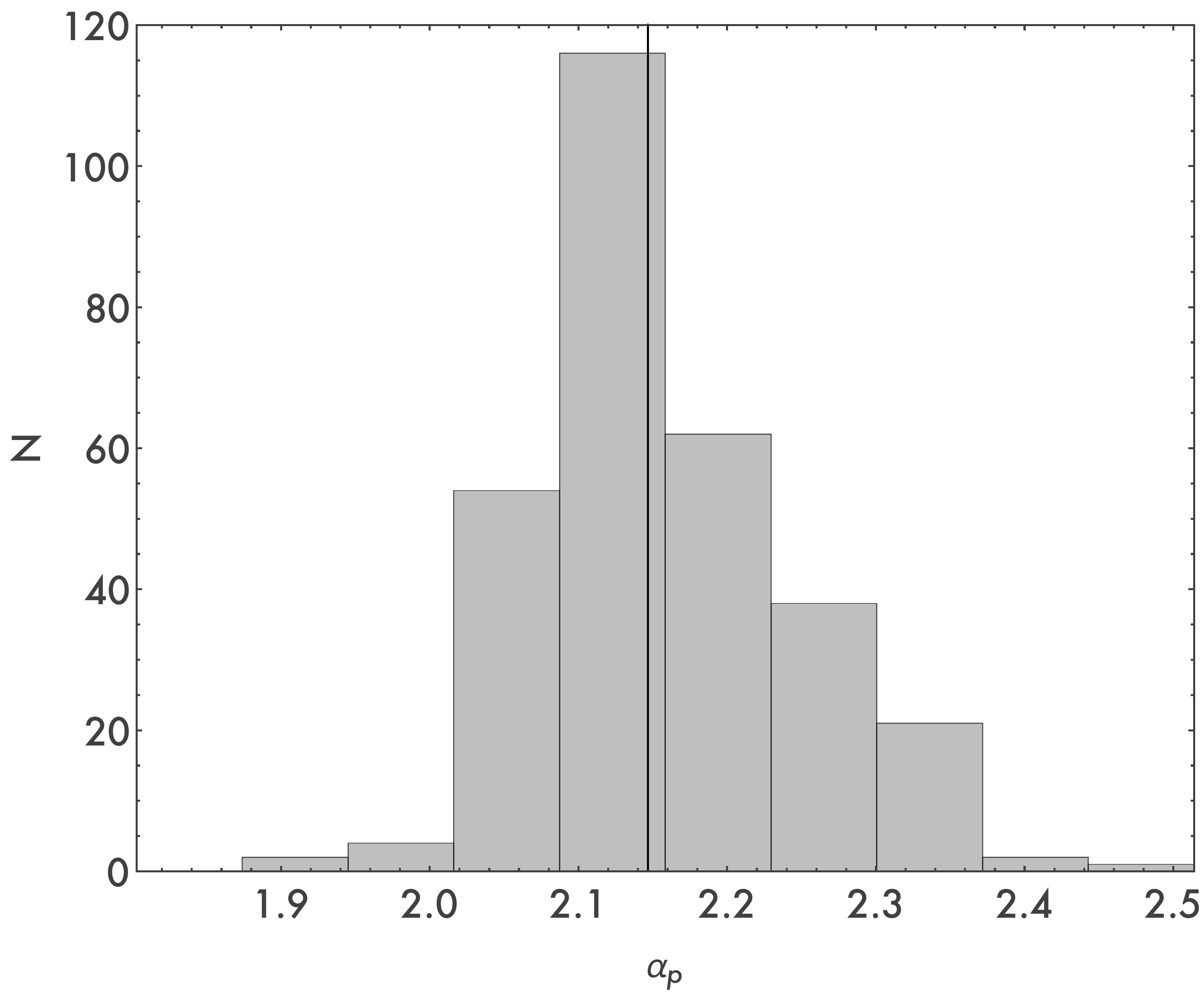} }}
\figcaption{Distributions of pitch angles resulting at $\xi = 100$ for
  particles injected at $\xi_i = 200$, all with a critical pitch
  angle, into a turbulent magnetic field with the same profile as used
  in Experiment 1.  The vertical line represents the value of the
  pitch angle that particles would have in the absence of turbulence.
}
\end{figure} 

We conclude our analysis by considering the effectiveness of
cosmic-rays propagating inward from the T-Taurisphere at ionizing a
circumstellar disk. Toward that end, we adopt the following fiducial
values of the system parameters (e.g., \citealt{hartbook}): disk
radius $R_d=50$ AU, disk mass $M_d = 0.05 M_*$, and stellar mass $M_*
= M_\odot$.  We also adopt standard temperature, density, and surface
density profiles of the form 
\be
T(r) = T_d \left({R_d \over r}\right)^{1/2}\,, \qquad 
\rho(r,z) = \rho(r) e^{-{z^2/H^2}}\, \qquad{\rm and}\qquad
\Sigma(r) = \Sigma_c \left({R_d\over r}\right)^p\,. 
\ee
In this treatment, the scale height $H(r)$ is taken to be the product
of the sound speed $a_s = \sqrt{kT/(2 m_p)}$ and rotation rate
$\Omega(r) = \sqrt{G M_*/r^3}$. The temperature and surface
density scales are $T_d = 30$ K and $\Sigma_c = 14$ g  cm$^{-2}$, respectively, and
the power-law index of the surface density is set at $p = 3/2$.  With these
choices, the expression for the mid-plane density becomes 
\be
\rho_m (r) = 1.3\times 10^{-13} \,{\rm g}\,{\rm  cm}^{-3}\, 
\left({R_d\over r}\right)^{11/4}\,.
\ee
Since protons are attenuated exponentially with a characteristic
length of $\Sigma_0 = 69$ g cm$^{-2}$ (Umebayashi \& Nakano 1981),
cosmic-rays entering the outer edge of the disk would lose their
ionizing potential after traveling a distance $L \sim \Sigma_0 / \rho
(R_d) \sim 35$ AU - which for a spiral path, occurs within the first
inward spiral.  Disk ionization is thus expected to result primarily
from particles coming inward near the equatorial plane, but moving a
few scale-heights above, where the density is significantly lower than
at the mid-plane. We note, however, that at large column densities,
the ionization rate is not due to protons, but to secondary particles 
(electrons and positrons) produced by charged and neutral pion 
decay through proton-proton collisions. The attenuation column for ionization is
$115$ g cm$^{-2}$ between 100 and 500 g cm$^{-2}$ from the
surface, whereas the ionization attenuation length is 
96 g cm$^{-2}$ above 500 g cm$^{-2}$ (Umebayashi \& Nakano 1981).
In any case, although our analysis was performed
for particles moving along the equatorial mid-plane, the field
structure within the disk will be fairly uniform, and our results then
applicable throughout the disk.  We are therefore able to construct a working 
estimate for the effects of magnetic turbulence on the 
inward propagation of cosmic rays from the T-Tauiosphere 
inward to the circumstellar disk. Nonetheless, a detailed 
analysis of the overall ionization structure within T-Tauri 
disks is left for future work. A host of additional effects 
should be considered, including ion-neutral damping and its 
back-reaction on the regions of the disk where turbulence 
can be active (see the discussion of Section 6.2).

\section{Conclusion} 
\label{sec:conclude} 

Cosmic rays represent an important source of ionization for the disks
associated with young stars. These disk environments provide the sites
for planet formation and the ionization levels affect the process.
This paper considers the propagation of cosmic rays through spiral
magnetic fields, such as those produced by stellar winds, with a focus
on the parameter space applicable to T Tauri star/disk systems. In
this treatment, turbulent fluctuations to the magnetic field are
included, and their effects on the inward propagation of cosmic rays
are determined. Our specific results are summarized below (Section
\ref{sec:summary}) along with a discussion of their implications
(Section \ref{sec:discuss}).

\subsection{Summary of Results} 
\label{sec:summary} 

In order to study the propagation of cosmic rays through turbulent
magnetic fields, we first construct a new coordinate system where one
of the coordinates follows the unperturbed magnetic field lines
(Section 2). The unperturbed field is assumed to follow a Parker
spiral. We then construct the perpendicular coordinates and the
necessary differential operators.  Note that T Tauri stars rotate
faster than the Sun, so that the winding parameters fall in a
different regime than for the Solar case. The new coordinate system
allows us to construct Alfv{\'e}nic field fluctuations (Section 3),
where the perturbations are perpendicular to the original field lines
and where the fluctuations manifestly have zero divergence. 

With the magnetic field structure and turbulent fluctuations
specified, we have carried out a large ensemble of numerical
integrations to follow the inward propagation of cosmic rays 
(Section 5).  The system is characterized by a number of parameters,
including the winding parameter of the original spiral structure 
$A=v_w/v_{rot}$, the amplitude of the turbulent fluctuations $\eta$,
the minimum wave number $k_1$, the spectral index of the turbulent
cascade $\Gamma$, the magnetic field strength of the stellar surface
$B_\ast$, the initial pitch angles $\alpha_i$, and the cosmic ray
energies as determined by the Lorentz factor $\gamma$. In addition,
due to the stochastic nature of the turbulence, each cosmic ray will
experience a different realization of the magnetic field fluctuations
as it propagates inward.

The results of our numerical experiments are presented in Figures 8 --
10 for different choices of the aforementioned system parameters.  In
the absence of turbulence, incoming cosmic rays with a given set of
initial conditions $(\alpha,\gamma)$ would reach a well-defined inner
turning point that represents an inner boundary to their sphere of
influence. The inclusion of turbulence replaces this well-defined
mirror point with a distribution of values. The width of this
distribution increases with the amplitude of the turbulent
fluctuations. In addition, for small amplitudes $\eta\ll1$, these
distributions can be modeled as Gaussians, where the turning point for the
non-turbulent fields lies at the center of the distribution. For
larger fluctuation amplitudes, however, the distributions become
highly non-gaussian, and the mean of the distribution falls outside
the turning point of the unperturbed system.  As a result, the
presence of turbulence acts to significantly reduce the flux of cosmic
rays that enter the inner parts of disks. The reduction becomes
significant, with the mean turning point radius becoming twice as
large, when the turbulent amplitude $\eta\simgreat0.3$. 

\subsection{Discussion} 
\label{sec:discuss} 

This paper has presented an semi-analytic treatment of cosmic ray
propagation through the turbulent magnetic fields expected in the
environments associated with young star/disk systems. As outlined
above, this work elucidates the basic physics of cosmic ray transport
and shows that turbulence acts to suppress the cosmic ray flux and
hence the ionization rate.  Although this subject is well developed
for the Solar Wind (e.g., see the reviews of \citealt{goldstein,tu}),
relatively little work has been done for young stars. The necessity
for understanding cosmic ray ionization rates was emphasized by
\cite{gammie}, who showed that cosmic rays cannot penetrate to the
mid-plane of circumstellar disks, so that dead zones with little
ionization develop. Additional suppression of the external cosmic ray
flux through the action of T Tauri winds was considered by
\cite{cleevestts}, who introduced the concept of a T Tauriosphere, a
region surrounding the star/disk system with suppressed cosmic ray
flux.  This paper studies the propagation of cosmic rays inward from
the T Tauriopause, toward the inner disk regions that are susceptible
to dead zones, and thus acts to connect these previous calculations.
Nonetheless, many avenues for additional work should be pursued.

This work focuses on the case where the unperturbed fields follow a
spiral form (as in \citealt{parker}). However, alternate --- and more
complicated --- field geometries should be explored. In addition, the
parameter space available for these star/disk/magnetic systems is
quite large and a full exploration of the entire
$(A,\alpha_i,k_1,\Gamma,B_\ast,\gamma)$ space is beyond the scope of
any single contribution. This work uses a simple prescription for the
turbulence, but more sophisticated treatments should be employed.
Finally, this work proceeds using semi-analytic methods. Since much of
the previous literature for circumstellar disks simply uses the
standard (single) value of the interstellar cosmic rate ionization
rate ($\zeta\sim10^{-17}$ s$^{-1}$, \citealt{umenakano}), this work
provides an important step forward. Nonetheless, full MHD simulations
of the problem should also be carried out. These numerical treatments
can include additional effects beyond what is possible with a
semi-analytic approach.  For example, we have assumed that the
turbulence remains active throughout the region of cosmic ray
propagation. However, this assumption could be modified by ion-neutral
damping, which reduces the amplitude of magnetic turbulence when the
frequency of magnetic waves is comparable to the frequency of
ion–neutral collisions. With effective damping, the turbulence levels
could have spatial dependence that is not addressed herein.

The results of this work inform a number of applications. Turbulence
in the disk itself is thought to provide one source for disk
viscosity, which is necessary for disk accretion to occur. The disk
turbulence is (most likely) driven by MHD instabilities such as MRI
\cite{balbus}, which requires ionization levels high enough to couple
the gas to the magnetic field. Cosmic rays provide an important source
of ionization, especially in the outer regions of the disk (beyond
$\sim10$ AU). This paper shows that the flux of cosmic rays can be
suppressed as the particles propagate inward from the Tauriopause to
the disk itself. This reduced cosmic ray flux will, in turn, reduce
the regions of the disk that is sufficiently ionized for the MRI to be
active.  An important topic for the future is to build models that
study cosmic ray propagation inward through both stellar winds and
magnetic turbulence.  The stellar winds suppress the cosmic ray flux
within their sphere of influence --- the T Tauriosphere --- which
specifies the inner boundary condition for the continued inward
propagation of cosmic rays as considered in this paper. The mirroring
effects, enhanced by turbulence, then provide additional suppression
of the flux. A combined treatment should thus study how the cosmic ray
flux in jointly modulated by the action of both stellar winds and
turbulent magnetic fields.

\acknowledgements

We would like to thank the referee for their careful reading of the manuscript 
and for their numerous and helpful comments.  We would also like to 
thank Ted Bergin and Ilse Cleeves for useful
conversations. M.F. thanks the Hauck Foundation and Xavier University
for funding support.  F.C.A. acknowledges support from the NASA
Exoplanets Research Program NNX16AB47G and from the University of
Michigan. 

\clearpage
\appendix

\section{Construction of Orthogonal Coordinates} 

In this Appendix, we show how the co-ordinate $p(\xi,\phi)$ can be
obtained from the co-ordinate $q$ using the conditions that these
co-ordinates are orthogonal.  For simplicity, we adopt a dimensionless
spherical co-ordinate system ($\xi, \theta,\phi$).  Starting with the 
definition 
\be
q \equiv \xi - 1 - \ln{\xi} - A \phi  \sin \theta_0 \,,
\ee
we now look for a co-ordinate $p$ such that 
$\nabla p \perp \nabla q$.
Since
\be
\nabla q = {\xi-1\over\xi}\hat\xi - {A\sin\theta_0\over\xi \sin\theta } \hat\phi \,,
\ee
this condition is met so long as
\be
\nabla p = \left[{A \sin\theta_0 \over \sin\theta} \hat \xi + \left(\xi-1\right)\hat\phi\right]\,f(\xi,\theta,\phi)\,,
\ee
which in turn requires
\be
{\partial p\over \partial \xi} = {A \sin\theta_0 \over \sin\theta} f(\xi,\theta,\phi)\,,
\ee
and
\be
{\partial p \over \partial \phi} = \xi (\xi-1) \sin\theta  f(\xi,\theta,\phi)\,.
\ee
Combining these conditions, one finds that 
$f(\xi,\theta,\phi)$ must satisfy the equation
\be
{A \sin\theta_0 \over \sin\theta} {\partial f \over \partial \phi} = \left(2\xi -1 \right) \sin\theta f 
+ \xi\left(\xi - 1\right)\sin\theta {\partial f\over\partial \xi}\,.
\ee
Here we use separation of variables and constrain our solutions to lie
on the surface given by $\theta = \theta_0$, and thus look for
solutions of the form
\be
f(\xi,\phi) = \Xi[\xi]\,\Phi[\phi]\,,
\ee
which yields the equation
\be
{A\over \sin\theta_0} {1\over \Phi} {d \Phi \over d\phi} = 
(2\xi-1) + \xi (\xi-1)  {1\over \Xi} {d\Xi\over d\xi}\,.
\ee
Upon substitution of the trivial solution
\be
\Phi = e^{m\phi}\,
\ee
one then is left to solve the equation
\be
{d\Xi \over \Xi} = \left[ {mA\over \sin\theta_0} - (2\xi -1) \right]\,{d\xi\over \xi(\xi-1)}\,.
\ee
Integrating this equation then yields
\be
\ln \Xi = (mA' -1)\ln (\xi-1) - (m A' +1) \ln \xi\,,
\ee
where $A' \equiv A/\sin\theta_0$.
In turn
\be
f(\xi,\phi) = \left[{(\xi-1)^{m A' -1} \over \xi^{m A' +1}}\right] e^{m\phi}\,,
\ee
and subsequently,
\be
p(\xi,\phi) = \left[{(\xi-1)^{m A' } \over \xi^{m A' }}\right] {e^{m\phi} \over m}\,.
\ee
The exponential dependence on $\phi$ in equation (A13) does not seem physical.  We note, however, that in the limit 
$m\rightarrow 0$, $f\rightarrow  \xi^{-1} (\xi-1)^{-1}$.
Equation (A4) can then be integrated to yield
\be
p = A \log\left({\xi-1\over \xi}\right)\,,
\ee
and
equation (A5) can be integrated to yield
\be
p = \phi \sin\theta_0\,.
\ee
Adding these solutions then yields the expression adopted in our work:
\be
p = A \ln \left(1-{1\over\xi}\right) + \phi  \sin\theta_0\,.
\ee

\section{Drift for Relativistic Particles}  

Following the analysis of Aguirre et al. (2010), we write the particle
velocity and acceleration in cylindrical co-ordinates in dimensionless
form: 
\be
{\vec \beta} =  \dot {\bar r} \hat r+ \bar r \dot\phi \hat\phi +\dot {\bar z}  \hat z \,,
\ee
and
\be
\dot {\vec\beta} = ( \ddot {\bar r}  - \bar r \dot \phi^2)\hat r +  
(\bar r \ddot\phi + 2  \dot {\bar r} \dot\phi)\hat\phi
+ \ddot {\bar z}\hat z \,,
\ee
and rewrite equation (32)  as
\be
( \ddot {\bar r}  -\bar r \dot \phi^2) = - {\dot {\bar z} R_*  \over \gamma A R_0 \bar r}
\qquad\qquad (\bar r \ddot\phi + 2  \dot {\bar r} \dot\phi) = 0
\qquad\qquad \ddot { \bar z}= {\dot {\bar r} R_*  \over\gamma A R_0 \bar r}\,.
\ee
The second and third equations can be integrated to yield constants of the motion:
\be
L = \bar r^2 \dot\phi\,,
\ee
and
\be
P = \dot {\bar z} - {R_* \over A\gamma R_0} \ln \bar r\,.
\ee 
Without loss of generality, we can set an initial condition for a particle with Lorentz factor
$\gamma$ at $\bar r = \bar r_i$ with  $\dot {\bar z}_i=0$, where $\bar r_i$ effectively replaces 
the constant $P$ in equation (B5):
\be
\dot {\bar z} = {R_* \over A\gamma R_0} \ln\left[{\bar r \over\bar r_i}\right]\,,
\ee
Integration then yields
\be
\bar z(\bar t)= \bar z_0 + {R_* \over A \gamma R_0 } \int_0^{\bar t}
 \ln\left[{\bar r(\bar t') \over\bar r_i}\right] d\bar t' \,.
\ee
Since $r_i$ is a periodic function with some periodicity $T_0$, the integral can be
written as 
\be
 \int_0^{\bar t}
 \ln\left[{\bar r(\bar t') \over\bar r_i}\right] d\bar t' = {\bar t \over T_0} \int_0^{T_0} \ln\left[{\bar r(\bar t') \over\bar r_i}\right] d\bar t' + G(\bar t)\,,
 \ee
where $G(\bar t)$ is a function with periodicity $T_0$.
Using equation (B3), one then obtains
\be
\bar z(\bar t)= \bar z_0 + {\bar t \over T_0} {A\gamma R_0 \over R_*}\, \int_{t_-}^{t_+}\left[{L^2 \over \bar r^2}
 + \dot {\bar r}^2\right] d\bar t' +  {R_* \over A \gamma R_0 } G(\bar t)\,,
\ee
 where $t_{-}$ and $t_+$ are the time values
where  $\bar r(\bar t)$ reaches its minimum and maximum values, and $L = {\bar r_i}^2 \dot \phi_i$.
The particle therefore drifts in the $z$ direction with a (dimensionless) drift speed
\be
v_{drift} = {A\gamma R_0 \over T_0 R_*}\, \int_{t_-}^{t_+}\left[{L^2 \over \bar r^2}
 + \dot {\bar r}^2\right] d\bar t'\,,
\ee
which is proportional to the Lorentz factor $\gamma$.

\newpage
$\,$

\end{document}